\documentclass[10pt,journal,compsoc]{IEEEtran}

\ifCLASSOPTIONcompsoc
  \usepackage[nocompress]{cite}
\else
  \usepackage{cite}
\fi

\usepackage{comment}

\ifCLASSINFOpdf
  \usepackage[pdftex]{graphicx}
\else
\fi

\usepackage{amsmath}
\usepackage{bm}
\usepackage{amsfonts}
\usepackage{multirow}
\usepackage{array}
\usepackage{xcolor}
\usepackage{pifont}
\newcommand{\cmark}{\ding{51}}%
\newcommand{\xmark}{\ding{55}}%

\begin{document}

\title{From the Lab to the Wild:\\Affect Modeling via Privileged Information}

\author{Konstantinos Makantasis, Kosmas Pinitas, Antonios Liapis,~\IEEEmembership{Member,~IEEE,}
        and Georgios N. Yannakakis,~\IEEEmembership{Member,~IEEE}
\IEEEcompsocitemizethanks{\IEEEcompsocthanksitem Konstantinos Makantasis is with the Department of Artificial Intelligence, Univerisity of Malta, Msida, Malta, MSD2080.\protect\\
\IEEEcompsocthanksitem Kosmas Pinitas, Antonios Liapis and Georgios N. Yannakakis are with the Institute of Digital Games, Univerisity of Malta, Msida, Malta, MSD2080.\protect\\
}
\thanks{Manuscript received Month Day, Year; revised Month Day, Year.}}

\markboth{IEEE Transactions on Affective Computing}%
{Makantasis \MakeLowercase{\textit{et al.}}: From the Lab to the Wild: Affect Modeling via Privileged Information}

\IEEEtitleabstractindextext{

\begin{abstract}
How can we reliably transfer affect models trained in controlled laboratory conditions (\emph{in-vitro}) to uncontrolled real-world settings (\emph{in-vivo})? The information gap between in-vitro and in-vivo applications defines a core challenge of affective computing. This gap is caused by limitations related to affect sensing including intrusiveness, hardware malfunctions and availability of sensors. As a response to these limitations, we introduce the concept of privileged information for operating affect models in real-world scenarios (in the wild). Privileged information enables affect models to be trained across multiple modalities available in a lab, and ignore, without significant performance drops, those modalities that are not available when they operate in the wild. Our approach is tested in two multimodal affect databases one of which is designed for testing models of affect in the wild. By training our affect models using all modalities and then using solely raw footage frames for testing the models, we reach the performance of models that fuse all available modalities for both training and testing. The results are robust across both classification and regression affect modeling tasks which are dominant paradigms in affective computing. Our findings make a decisive step towards realizing affect interaction in the wild. 
\end{abstract}

}

\maketitle

\IEEEdisplaynontitleabstractindextext

\IEEEpeerreviewmaketitle

\IEEEraisesectionheading{\section{Introduction}\label{sec:introduction}}

\IEEEPARstart{D}{espite} the recent advances in Affective Computing (AC), largely based on today's powerful deep learning algorithms \cite{toisoul_estimation_2021,botelho_improving_2017,li_how_2020,trigeorgis_adieu_2016,tzirakis_end--end_2021,makantasis2021pixels}, affect modeling approaches are still struggling to reliably transfer models trained on data collected in the laboratory (\emph{in-vitro}) to real-world settings (\emph{in-vivo}), that is, in the wild. Sensing affect in controlled laboratory conditions results in high-quality data characterized by precise multimodal measurements. On the contrary, the quality of affect measurements in uncontrolled, real-world settings is limited by a number of environmental and experimental conditions. Consequently, the information gap between in-vitro and in-vivo affect modeling limits the transferability of affect models to real-world applications.  

Compared to in-vitro, the quality of in-vivo affect measurements is limited by several factors. First, the environment affects the sensing equipment (e.g. cameras, microphones, physiology sensors). Therefore, the acquired measurements are likely to be either corrupted by experimental noise due to sensors' failures or biased due to environmental conditions under uncontrolled settings. As a result the presence of noise and bias deteriorates the performance of affect models. Second, exploiting multimodal information to model affect is the norm in current AC approaches. However, as AC occurs in-vivo, sensors for capturing multimodal information are unavailable (e.g. electroencephalogram measurements). We cannot assume---and arguably we should not assume---that users engaged in affect interaction will have access to a plethora of specialized sensors for measuring affect in their houses, their cars \cite{eyben2010emotion,braun2019improving}, or public spaces including museums \cite{kim2015measuring}, hospitals and rehabilitation centers \cite{tripathi2021advancing}. Interestingly, such real-world settings define some of the most popular application domains of AC. Finally, capturing information about users in real-world settings comes with a cost in terms of intrusiveness (e.g. requiring users to install and use sensors properly) and privacy (e.g., access to sensitive information via smartphone's webcam and microphone). 

This study aims to overcome the current limitations of affect sensing in the wild by introducing the notion of privileged information to affect modeling. In particular, the Learning Using Privileged Information (LUPI) paradigm \cite{vapnik2009new, vapnik2015learning} is best suited for tasks that present different amounts of information available during the models' training and testing phases. In the context of AC privileged information is of utmost importance when there is an information gap between the development and the deployment of an affect model. Our hypothesis is that the LUPI paradigm can be beneficial for mitigating the limitations of affect sensing in the wild, leading to affect models that achieve similar performance in-vitro and in-vivo conditions. We test our hypothesis using the popular RECOLA and SEWA affect databases \cite{ringeval2013introducing,kossaifi2019sewa}. These databases consist of multimodal user information annotated across two affect dimensions; arousal and valence. User measurements include facial images, visual and audio features, and electrodermal activity (EDA) and electrocardiogram (ECG) biosignals. They also include continuous arousal and valence traces performed by six annotators. We consider the information from facial images available both in-vitro and in-vivo as capturing facial images requires no special equipment, just conventional cameras. On the contrary, we consider all other modalities to be available only in-vitro since capturing this information requires specialized laboratory equipment or video and audio processing algorithms.

Similarly to \cite{makantasis2019pixels,makantasis2021pixels}, our models of affect rely on raw images both for training and testing phases. Additional knowledge from privileged information is injected into the models during training. During testing, however, the developed models make predictions using only raw images without dependence on privileged information, as would be the case of an in-vivo setting. Our experimental outcomes suggest that exploiting privileged information yields models that perform consistently better than models trained only with images. More importantly, in many cases, models trained under the LUPI paradigm perform equally well as the models that base their predictions on fusing all available modalities (images and privileged information). The results obtained are consistent across the two affect dimensions of arousal and valence and robust across the two most dominant learning paradigms used in AC: classification and regression.  This suggests that our approach is robust for training models that can perform in the wild without extensive hyperparameter tuning. Our findings indicate that privileged information is a critical milestone for bridging the gap between in-vitro and in-vivo affect modeling and realising AC in the wild. Models trained on information only available in a lab setting can perform equally well when that information is unavailable or distorted in the wild. Following LUPI, models of affect gain on robustness, unobtrusiveness, accessibility and practicality. 

This paper builds upon and extends significantly our earlier work \cite{makantasis2021privileged} in several ways. First, our study in \cite{makantasis2021privileged} applies the LUPI paradigm to affect classification. The present study goes one step further and applies the LUPI paradigm to affect modelling problems treated as both classification and regression, covering the vast majority of affect modeling tasks. Second, in \cite{makantasis2021privileged} privileged information is injected into affect models at the output layer. This study proposes a general methodology for injecting privileged information at any hidden layer of deep learning-based affect models. Therefore, the proposed methodology can derive affect models that exploit privileged information irrespective of the models' architecture choices. Third, in this study, we go beyond arousal prediction within the domain of digital games \cite{makantasis2021privileged,makantasis2021pixels,makantasis2019pixels}, and evaluate the proposed scheme in two popular publicly available affect databases across two affect dimensions: valence and arousal. We should also mention that one of these databases has been developed for testing the performance of affect models in the wild. The evaluation results suggest that exploiting privileged information can boost the performance of affect models that operate in the wild.

\section{Related Work}
This section covers the related areas of pixel-based affect modeling, multimodal affect modeling based on audiovisual information and physiology measurements, affect modeling in the wild, and affect modelling under missing data or modalities. 

\subsection{Pixel-based and Multimodal Affect Modeling}

Due to the richness of information encoded in videos and images, eliciting and modeling emotion via visual cues has been a core interest in affective computing \cite{picard2000affective}. Before the deep learning era, the dominant approach for representing visual content was based on the construction of ad-hoc handcrafted features. Along this line, sophisticated visual descriptors, such as scale-invariant feature transform \cite{lowe2004distinctive}, histogram of oriented gradients \cite{dalal2005histograms} and linear binary patterns \cite{ahonen2004face}, have been widely used to produce high-level representations of facial patterns used for recognizing emotions \cite{ko2018brief,zheng2010emotion,julina2019facial,dahmane2011emotion}. Emotion recognition approaches based on handcrafted visual features are characterized by low computational and memory requirements, and thus, are still being studied for use in real-time embedded systems \cite{suk2014real}.  
In the last decade, convolutional neural networks (CNN) led to a breakthrough in computer vision and visual information processing. During their training, CNNs automatically produce high-level representation of visual information, eliminating the need for handcrafted features construction. Baveye \textit{et al.} \cite{baveye2015deep} proposed one of the first approaches using CNNs for predicting dimensional affective scores from videos. However, the limited number of training data prevented the learning of data-hungry CNN models.

The development of medium- and large-scale affect corpora \cite{zhang2012finding,zafeiriou2017aff} established the use of deep learning in affect modeling \cite{martinez2013learning}. Breuer and Kimmer \cite{breuer2017deep} demonstrated the capacity of CNNs to jointly learn various facial expression recognition tasks. Jung \textit{et al.} \cite{jung2015joint} boosted the performance of facial-based affect models by exploiting high-level spatiotemporal representations of facial action points produced by CNNs. Ng \textit{et al.} \cite{ng2015deep} proposed transfer of learning across CNNs for emotion recognition through visual cues. Based on the hypothesis that gameplay footage embeds players' affect, studies in \cite{makantasis2019pixels,makantasis2021pixels} used CNNs to map raw gameplay footage to players' arousal. Finally, Martinez \textit{et al.} \cite{martinez2013learning} presented the first application of CNN models for detecting affect via physiological signals such as skin conductance. 

Along with visual information, additional modalities such as audio and physiology measurements have been used for modeling affect. The hypothesis that additional modalities can reveal different aspects of emotions triggered the collection of multimodal information datasets such as the RECOLA database \cite{ringeval2013introducing} used in this study, the DEAP \cite{koelstra2011deap}, AMIGOS \cite{correa2018amigos} and SEMAINE \cite{mckeown2011semaine} datasets. The dominant approach for processing multimodal information is based on fusing the different modalities into a common representation, which is then used as input to machine learning models. Indicatively, Tzirakis \textit{et al.} \cite{tzirakis2017end} propose a CNN and a deep residual network to combine auditory and visual information for emotion recognition. Ranganathan \textit{et al.} \cite{ranganathan2016multimodal} used deep belief networks to generate multimodal features from face, body gesture, voice and physiology measurements in an unsupervised manner. Siriwardhana \textit{et al.} \cite{siriwardhana2020multimodal} used self-supervised learning to fuse text, audio and visual information along with transformers for recognizing affect. Seng and Ang \cite{seng2019multimodal} presented emotion modeling techniques based on mulitmodal unstructured big-data, while Abdullah \textit{et al.} \cite{abdullah2021multimodal} presented a survey on the application of deep learning models to multimodal emotion recognition.

The methods discussed above model affect based on visual only or multimodal information captured in well-defined and controlled laboratory conditions. Collecting, however, data in a laboratory requires specialized hardware and software that might not be available in the wild, and yields noise-free and unbiased datasets. Both of these characteristics limit the application of affect models trained with laboratory-generated data to real-world settings. This paper aims to take affective modeling outside of a laboratory's closed boundaries by building models able to predict affect using information that is available in the wild, and at the same time, exploit knowledge through privileged laboratory-generated information.

\subsection{Affect Modeling In The Wild}

Affect modeling in the wild focuses on developing models able to analyze the emotional state of humans in real-life scenarios that entail uncontrolled conditions. To mitigate the problem of noisy, distorted and biased data, large databases \cite{zafeiriou2017aff,kollias2019expression,mavadati2013disfa,zhang2014bp4d,zheng2018multimodal,kossaifi2019sewa} that simulate human emotions in the wild are necessary \cite{kollias2016line}. The availability of large affect corpora have enabled powerful deep learning models that achieve state-of-the-art affect modeling results. 

AffWildNet proposed by Kollias \textit{et al.} \cite{kollias2017recognition,kollias2019deep} achieved the best performance in the \emph{Aff-Wild} challenge \cite{zafeiriou2017aff} by combining CNNs and recurrent neural networks to accurately capture face dynamics. The EmoFAN deep learning model \cite{toisoul_estimation_2021} jointly predicted discrete emotional states and continuous affect dimensions by building upon the face alignment network proposed in \cite{bulat2017far}, thereby achieving the best performance on the AfewVA dataset \cite{kossaifi2017afew}. Aspandi \textit{et al.} \cite{aspandi2020adversarial} estimated affect in the wild by exploiting adversarial neural networks that build high-level representations of audiovisual information. Parthasarathy and Sundaram \cite{parthasarathy2021detecting} demonstrated that multimodal deep learning affect models can significantly improve affect detection in the wild. They use multimodal transformers to capture and exploit temporal dynamics of audiovisual information towards detecting affect states. Finally, Kollias and Zafeiriou \cite{kollias2021affect} proposed a unified framework for affect modeling in the wild that considers facial expressions and categorical affect, facial action units, and dimensional affect representations. 

Although the studies listed above model affect in the wild, they all require large affect corpora to reduce the impact of noise and bias on the performance of affect models. Instead, this paper relies on the use of privileged information to effectively train and reliably transfer models of affect from controlled laboratory conditions to real-world settings. Rather than use training data captured in the wild, we use high quality laboratory measurements as privileged information for assisting the training of the models, which can then be applied and operate in the wild. Although exploiting privileged information has been proposed in \cite{makantasis2021privileged} for modeling players' arousal within the domain of digital games, this study extends the aforementioned work by applying the LUPI paradigm to two popular affect datasets beyond games for modeling both arousal and valence.

\subsection{Affect Modelling under Missing Data/Modalities}
Affect modelling in the wild may suffer from missing or corrupted data due to unforeseen sensor malfunctions. The study in \cite{du2018semi} proposes a semi-supervised multi-view model to address the problem of missing modalities. Their approach treats a missing modality as a latent variable which is integrated out during inference. The works in \cite{fortin2019multimodal,page2019multimodal} formulate emotion recognition as a multitask learning problem and leverage several classifiers under all combinations of different modalities to avoid the missing data/modality problem. The authors in \cite{zhao2021missing} propose the learning of joint multimodal representations able to predict the representations of any missing modality. They apply their methodology for trimodal emotion recognition using visual, acoustic and textual information. The study in \cite{parthasarathy2020training} investigates the performance of state-of-the-art transformers for bimodal emotion recognition problems where one of two modalities is missing. The authors of \cite{wang2022m2r2} propose a framework based on iterative data augmentations to address the problem of multimodal emotion recognition in conversation tasks with missing modalities. The aforementioned studies attempt to recover information about the missing data or modalities and exploit that information during a model's training or inference. Our approach, instead, does not target the problem of missing data or modalities. It is based on privileged information that is able to transfer knowledge encoded in models trained using a large set of information modalities to models trained on a smaller set of modalities.

Another approach to deal with missing modalities is based on the idea of dynamic fusion. The works in \cite{chen2019selective,yang2020adaptive} automatically estimate the importance of each modality during training and weigh it accordingly. During testing they exploit the learnt weighting scheme to fuse different modalities based on given inputs. These approaches are different than ours in the sense that our model is trained on several modalities and during testing it operates on a subset of them. Our aim is to introduce to a model knowledge from modalities that are not available during testing and not to fuse several modalities given the current inputs.

Finally, the study in \cite{abavisani2019improving} proposes to use separate networks per available modality and then to enforce them to collaborate and learn common semantics across modalities. During testing only the networks that correspond to the available modalities are used. This approach is also different than ours. First, it focuses on correlation matrices of the latent representations to learn common semantics. Second, it employs several unimodal classifiers and a knowledge transfer scheme from all classifiers to the one that will operate with missing modalities. Our approach uses just two classifers, one teacher and one student, and it is not based on correlation-based knowledge transfer. Instead, it uses the knowledge from the teacher to guide, via modifying the loss function, the learning of the student.

\section{Use Cases and Data Preprocessing}  
This section presents the datasets used for experimentally validating our proposed methodology and the data preprocessing steps.

\subsection{Datasets}
To test the impact of privileged information on affect modeling and investigate the degree to which it can transfer knowledge from in-vitro experiments to in-vivo AC applications we use two multimodal databases: RECOLA \cite{ringeval2013introducing} and SEWA \cite{kossaifi2019sewa}. 

RECOLA consists of audio, visual and physiological (EDA and ECG) recordings of online dyadic interactions between 34 participants. Since RECOLA has been used for audiovisual emotion recognition challenges, the creators split the database into two parts; data form 23 participants used for training and developing models of affect are publicly available, while the rest serve for evaluating the performance of the developed models and are not (and will not be) made publicly available. Six assistants (three males and three females) annotated the collected data in terms of arousal and valence. The annotations are continuous, bounded in the range of $[-1,1]$ and provided at 25Hz. 

SEWA contains recordings of 398 volunteers watching various advertisements and discussing them via a video-chat software. Volunteers' behaviour captured in completely unconstrained, real-world environments using webcams and microphones. Five raters provided continuous valence and arousal labels for the audio-visual recordings at $\sim$66Hz. These annotations were combined to a single ground truth that maximally correlates annotations from all raters. Finally, the ground-truth was normalised in $[0,1]$. In this study, we use the SEWA basic dataset, which consists of 538 short (10-30 second long) segments cut from the full video-chat recordings. For the rest of the paper, we refer to the SEWA basic dataset as SEWA. 

Along with the raw recordings, the RECOLA database creators provide features that describe each information modality. For audio information in RECOLA, besides raw audio files, probability of voice activity detection and eGeMAPS acoustic features \cite{eyben2015geneva} are also provided. Features describing the statistical properties of EDA and ECG are also given. Finally, visual information is described by raw $1080 \times 720$ video frames, probability of face detection, optical flow, and detection and movement of 15 emotion-related facial action units. In SEWA, we created 65 OpenSMILE \cite{eyben2013recent} acoustic features from the audio recordings. For RECOLA, we consider the video frames as \emph{pixel information} and the remaining metrics as \emph{visual features} which require complex software to process that may not be available in the wild.

\begin{table}[t]
\centering
\caption{Information modalities in RECOLA and SEWA within each time window along with their dimension. We treat only pixel information as non-privileged.} 
\begin{tabular}{p{6cm}|p{1.8cm}} 
\textbf{Modality -- RECOLA} & \textbf{Dimension} \\  \hline \hline
Pixel Information & 320 $\times$ 180 $\times$ 5 per second \\ \hline
Audio Features (e.g. eGeMAPS, voice activity) & 131\\ \hline
Visual Features (e.g. facial action units) & 41\\ \hline
Electrocardiogram (ECG) Features & 54\\ \hline
Electrodermal activity (EDA) Features & 63 \\ 

\multicolumn{2}{c}{} \\

\textbf{Modality -- SEWA} & \textbf{Dimension} \\  \hline \hline
Pixel Information & 320 $\times$ 180 $\times$ 5 per second \\ \hline
Audio Features (e.g. OpenSMILE features) & 65\\ 
\end{tabular}
\label{table:modalities}
\end{table}

\subsection{Data Preprocessing}

This study aims to produce models of affect that predict arousal and valence based on different information modalities. We split the interaction session of each participant using overlapping windows. The sliding step and the length of the windows are hyperparameters we consider. In this study, we conduct experiments for a sliding step of 400ms and window lengths of 1, 2 and 3 seconds. Using a fixed sliding step and overlapping time windows, the dataset size (number of time windows) is not affected by the windows' length. By varying the windows' length, the amount of temporal information encoded in each window changes affecting both audiovisual information and physiology features. 

After splitting the multimodal dataset across time windows, the information associated with each window corresponds to a sequence of raw footage frames concatenated along the channels dimension and the mean values of audiovisual and physiology features. To reduce the computational cost, we use grayscale footage frames resized to $320 \times 180$ pixels and frame skipping of five frames. Regarding RECOLA, as the data annotation is conducted by six assistants, we use the median annotation values per time instance to mitigate annotators' disagreement \cite{grewe2007emotions}. For SEWA, we use the audiovisual annotation values provided with the dataset. Then, the arousal and valence ground truth labels for each window correspond to the mean annotation values within the window's duration \cite{makantasis2021pixels,melhart2021again}. Table \ref{table:modalities} summarizes the information modalities that describe each of the time windows.

\section{Affect Modeling Using Privileged Information}

In this section we detail the \emph{Learning Using Privileged Information} paradigm \cite{vapnik2009new} for building models of affect capable of generalizing in the wild, as well as the architecture of the employed machine learning models.

\subsection{Learning Using Privileged Information}

Learning Using Privileged Information (LUPI) \cite{vapnik2009new, vapnik2015learning} addresses problems characterized by an asymmetric distribution of information between training and test time; specifically, additional information is given about the training data, which is not available at test time. This setting is prevalent in affective computing. A plethora of different information modalities can be captured in controlled laboratory conditions using specialized hardware and software. In the wild, however, it is impossible to capture the same modalities due to sensors' cost, noisy environments, and invasive capturing procedures.
LUPI provides the means to \textit{transfer knowledge} from all the available modalities to a machine learning model that makes predictions using only a subset of these modalities \cite{sharmanska2013learning,lopez2016unifying}. In other words, LUPI allows an affect model to be trained exploiting knowledge that comes from all the modalities captured in a laboratory setting or via specialized software and hardware. During test time, however, the same model makes predictions using only those modalities that are available in the wild. The information that is not available during test time is called \emph{privileged information}.

As far as the RECOLA database is concerned, we treat as privileged the information that corresponds to physiology and ausiovisual features provided by the database creators (see Table \ref{table:modalities}). For SEWA, we consider audio features as privileged. Our choice is justified by the fact that capturing physiology requires specialized sensors, while constructing physiology and audiovisual features implies the employment of specific software algorithms. On the contrary, we consider information that comes from raw footage frames as non-privileged (captured using conventional cameras) being available both at training and test times. 

Below, we describe transferring knowledge from privileged information to a machine learning model. At this point, we should clarify that transferring knowledge using LUPI is different from the transfer of learning techniques used in deep learning \cite{ng2015deep}. Transfer of learning targets small-sample setting problems by finetuning a model trained for a specific task such that it performs well in a similar task. On the contrary, using LUPI focuses on problems with asymmetric distribution of training/testing information and trains the models from scratch. 

This study explores the use of privileged information with neural network-based models of affect. Before transferring knowledge that comes from privileged information, we first have to represent it appropriately. Following \cite{hinton2015distilling,lopez2016unifying,vapnik2017knowledge}, we represent that knowledge within the latent and the output representations of a neural network that has been trained and makes predictions based on all available modalities or on privileged information only. This model is called \textit{teacher}. Having a teacher model trained, we can transfer knowledge from privileged information to another model called \textit{student}. The transfer of knowledge can be achieved by feeding the model with only those modalities on information that is available in the wild and force it during training to balance between the learning task's loss and learning latent representations that match those of the teacher model. After training, the student model makes predictions based only on the information that is available in the wild, without any dependence on the teacher model or privileged information. 

To be more rigorous, let us denote as $\bm x$ and as $\bar{\bm x}$ the information describing a specific sample fed to a student and a teacher model, respectively, and as $y$ the sample's affect ground truth label. By denoting as $S_l(\bm x) \in \mathbb R^d$ and as $T_k(\bar{\bm x}) \in \mathbb R^d$, respectively, the latent representations at the $l$-th layer and the $k$-th layer of the student and the teacher models, then the loss that the student is minimizing during training can be defined as:
\begin{equation}
\label{eq:general_loss}
    L_{st} = (1 - \alpha){\cdot}L(S_o(\bm x), y) + \alpha{\cdot}D(S_l(\bm x), T_k(\bar{\bm x}))
\end{equation}
where $\alpha \in [0,1]$, $L$ stands for the learning task's loss (e.g. mean squared error for regression tasks or cross-entropy loss for classification tasks), $S_o(\bm x)$ is the output of the student model, and $D$ is a distance metric penalizing deviations between student's and teacher's latent representations.  

In the present study, we focus on regression and classification tasks since these are the two most dominant paradigms in affect modeling. In the case of classification, we develop models that predict high vs low arousal/valence while in regression, our models aim to predict the continuous ground truth label of affect.

For classification problems the student loss in Eq.~\eqref{eq:general_loss} is defined as:
\begin{equation}
\label{eq:class_loss}
    L_{st} = (1 - \alpha){\cdot}L_{CE}(S_o(\bm x), y) + \alpha{\cdot}L_{KL}(S_o(\bm x), T_o(\bar{\bm x}))
\end{equation}
where $T_o(\bar{\bm x})$ are the probabilistic predictions of the teacher model, $L_{CE}$ is the cross-entropy loss, and $L_{KL}$ is the Kullback-Leibler divergence between student's and teacher's probabilistic predictions. Kullback-Leibler divergence is a statistical distance measuring the difference between two distributions---in our case, the probability distributions over the available classes for the student and teacher models---and it has been used within the knowledge distillation paradigm for transferring knowledge between different models \cite{hinton2015distilling}. By examining the relation in Eq.~\eqref{eq:class_loss}, it can be made clear that in the case of classification, the student has to minimize the classification loss and at the same time follow the probabilistic predictions of the teacher. 

\begin{figure*}[!tb]
	\begin{minipage}{1.0\linewidth}
		\centering
		\centerline{\includegraphics[width=1.0\linewidth]{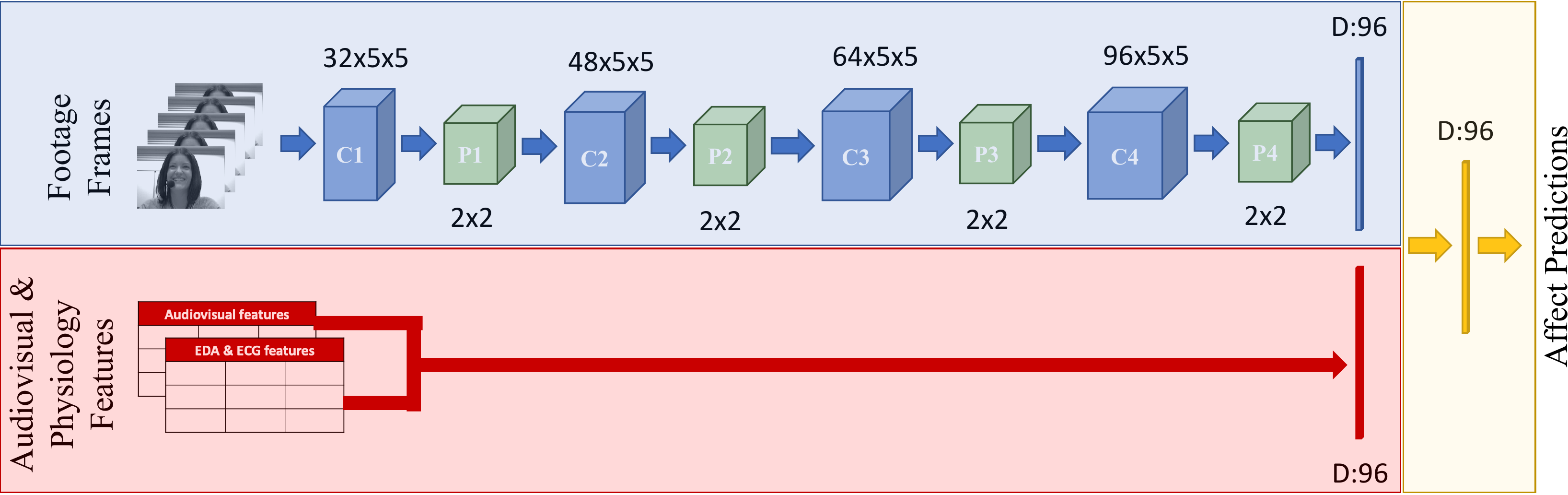}}
	\end{minipage} 
	\caption{Architecture of the employed models of affect. The convolutional, max-pooling and dense layers are demoted by ``C'', ``P'' and ``D'' respectively. The blue-shaded stream corresponds to the PixelNet and student models and the red-shaded to the PrivNet. The FusionNet combines the blue and red stream with the yellow-shaded module that fuses the information from the different modalities. }
	\label{fig:architecture}
\end{figure*}

For regression problems, the above procedure can not be followed. The output of regression models is a real value and not a probability distribution, and therefore Kullback-Leibler divergence cannot be used. In addition, under a regression setting, requiring from the student model to follow a teacher's predictions will force the student's output away from the desired ground truth labels. For all those reasons, we inject knowledge about privileged information at the penultimate layer of the student model \cite{vapnik2017knowledge}. To achieve that, we force the output of the penultimate layer of the student model to be close to the corresponding output of the teacher model by defining the student loss in Eq.~\eqref{eq:general_loss} as:
\begin{equation}
\label{eq:reg_loss}
    L_{st} = (1 - \alpha) L_{MSE}(S_o(\bm x), y) + \alpha L_{CS}(S_p(\bm x), T_p(\bar{\bm x}))
\end{equation}
where $L_{MSE}$ stands for the mean square error, $S_p(\bm x)$ and $T_p(\bar{\bm x})$ the output of the penultimate layer of the student and teacher models, and $L_{CS}$ the cosine similarity. The relation in Eq.~\eqref{eq:reg_loss} indicates that the student minimizes the regression loss and at the same time it tries to match the representation at its penultimate layer to the representation produced by the teacher. By forcing the latent representations of the student to match the latent representations of the teacher, knowledge about privileged information can be injected at \textit{any} layer of a neural network-based affect model. Therefore, the proposed approach can be applied to any deep learning affect model irrespective of the architecture design choices. 

Both Eq.~\eqref{eq:class_loss} and Eq.~\eqref{eq:reg_loss} rely heavily on the $\alpha$ parameter, which determines the impact of the privileged information on the training of the student model. By increasing the value of $\alpha$, we force the student model to weigh more the knowledge coming from the teacher models and pay less attention to ground truth labels. When $\alpha=0$ the student considers only ground truth labels without exploiting privileged information. On the contrary, when $\alpha=1$ the student follows the teacher disregarding any information from the ground truth labels.

In the presented case study, we employ two teacher models. The first model is trained using only privileged information, and the second using all available modalities (privileged information plus pixel information). The student model is trained using raw interaction footage frames and teacher's knowledge. We should emphasize that after training the student makes predictions using information \textit{solely} from raw frames. 

\subsection{Machine Learning Models of Affect}

\begin{table}[t]
\centering
\caption{Information modalities used for training and testing the different models. Details of each modality are found in Table \ref{table:modalities}.} 
\small
\begin{tabular}{l|c|c|c|c|c|c|c|c|c|c}
& \multicolumn{2}{c|}{\textbf{Pixel}} & \multicolumn{2}{c|}{\textbf{Audio}} & \multicolumn{2}{c|}{\textbf{Visual}} & \multicolumn{2}{c|}{\textbf{ECG}} & \multicolumn{2}{c}{\textbf{EDA}}  
\\ \cline{2-11}

\textbf{\underline{RECOLA}} & \rotatebox{90}{Train~} & \rotatebox{90}{Test} 
& \rotatebox{90}{Train~} & \rotatebox{90}{Test} 
& \rotatebox{90}{Train~} & \rotatebox{90}{Test} 
& \rotatebox{90}{Train~} & \rotatebox{90}{Test} 
& \rotatebox{90}{Train~} & \rotatebox{90}{Test} 
\\ \hline \hline
\textbf{PixelNet} & \cmark & \cmark & \xmark & \xmark & \xmark & \xmark & \xmark & \xmark & \xmark & \xmark \\ \hline
\textbf{FusionNet}& \cmark & \cmark & \cmark & \cmark & \cmark & \cmark & \cmark & \cmark & \cmark & \cmark \\ \hline
\textbf{PrivNet}& \xmark & \xmark & \cmark & \cmark & \cmark & \cmark & \cmark & \cmark & \cmark & \cmark \\ \hline
\textbf{StudentNet} & \cmark & \cmark & \cmark & \xmark & \cmark & \xmark & \cmark & \xmark & \cmark & \xmark \\

\end{tabular}

\vspace{0.1in}

\begin{tabular}{l|c|c|c|c|c|c|c|c|c|c}
& \multicolumn{2}{c|}{\textbf{Pixel}} & \multicolumn{2}{c|}{\textbf{Audio}} & \multicolumn{2}{c|}{\textbf{Visual}} & \multicolumn{2}{c|}{\textbf{ECG}} & \multicolumn{2}{c}{\textbf{EDA}}  
\\ \cline{2-11}

\textbf{\underline{SEWA}} & \rotatebox{90}{Train~} & \rotatebox{90}{Test} 
& \rotatebox{90}{Train~} & \rotatebox{90}{Test} 
& \rotatebox{90}{Train~} & \rotatebox{90}{Test} 
& \rotatebox{90}{Train~} & \rotatebox{90}{Test} 
& \rotatebox{90}{Train~} & \rotatebox{90}{Test} 
\\ \hline \hline
\textbf{PixelNet} & \cmark & \cmark & \xmark & \xmark & - & - & - & - & - & - \\ \hline
\textbf{FusionNet}& \cmark & \cmark & \cmark & \cmark & - & - & - & - & - & - \\ \hline
\textbf{PrivNet}& \xmark & \xmark & \cmark & \cmark & - & - & - & - & - & - \\ \hline
\textbf{StudentNet} & \cmark & \cmark & \cmark & \xmark & - & - & - & - & - & - \\

\end{tabular}
\label{table:models}
\end{table}

The student model---we name it \textit{StudentNet}---used in this study is a 2D CNN with four convolutional layers that receives as input a sequence of grayscale footage frames concatenated along the channels dimension. The first two convolutional layers consist of 32 and 48 learnable kernels of dimension 5 x 5 and stride equal to 2. The third and fourth convolutional layers consists of 64 and 96 learnable kernels of dimension 3 x 3 and stride 1. A $2 \times 2$ max-pooling layer follows each of the convolutional layers. The last convolutional layer's output is fed to a dense layer with 96 hidden neurons and then is propagated to the output layer. At the penultimate layer of the student model we also use dropout with 10\% probability.

As mentioned above, we use two teacher models; \textit{PrivNet} trained with only privileged information, and  \textit{FusionNet} trained with all information modalities (see Table \ref{table:models}). PrivNet is a fully connected feed forward neural network with one hidden layer with 96 neurons. FusionNet is a two-stream network: the stream that processes the footage frames has the same architecture as the student model, while the stream that processes the privileged information is similar to PrivNet. The outputs of the two streams are concatenated and pass through a dense layer with 96 neurons before they are fed to the output layer.

To evaluate the impact of privileged information on the affect model's performance, we build a baseline model: a CNN trained and making predictions by exploiting only pixel information of the footage frames. We name this model \textit{PixelNet} following the reported benefits of modeling affect solely from pixels \cite{makantasis2019pixels,makantasis2021pixels}. The architecture of PixelNet is the same as the architecture of the student model.

For all models, we use the Adam optimizer \cite{kingma2014adam} with learning rate $0.001$, batches of size 256 to train the models, and ReLU as activation function for all models' layers. Table \ref{table:models} presents the modalities used for training and testing the performance of the employed models. Figure \ref{fig:architecture} presents the architecture of all employed models. The blue-shaded stream corresponds to PixelNet and StudentNet and the red-shaded to PrivNet. FusionNet consists of both blue and red streams, and it employs the yellow-shaded module for fusing the information from the different modalities before the output layer. 

It should be noted that this study does not aim to produce state-of-the-art results. Therefore, we did not conduct any architecture search, including the investigation of different fusion strategies, for deriving the best possible learners for the problems at hand. This study aims, first, to rigorously formalise a methodology for exploiting privileged information in affect modelling and under different learning paradigms, and, second, to showcase the benefits of infusing privileged information into affect models.

\section{Results}

This section presents the framework for evaluating the impact of privileged information on affect modeling and the experimental results obtained.

\subsection{Evaluation Framework}
\label{ssec:evaluation_framework}

As mentioned earlier, we treat affect modeling both as a regression and as a classification task. In the former our models attempt to predict continuous ground truth affect annotations. In the latter, however, the models learn to map user information to low and high arousal/valence classes. Therefore in classification, we use a split criterion value $t$ that determines the class (low vs high) of each data point. We set the median values as our split criterion ($t=0.07$ and $t=0.05$ for arousal and valence, respectively) in an attempt to create balanced datasets for both affect dimensions. 
Having defined parameter $t$, we assign the examples whose annotation values is larger than $t+\epsilon$ to the high arousal/valence class, and the examples with annotation values smaller than $t-\epsilon$ to the low arousal/valence class. The $\epsilon$ parameter determines a region around the class-splitting value within which annotation values are treated as uncertain and ignored during affect classification to avoid unstable classifiers due to trivial differences in their inputs \cite{makantasis2019pixels,makantasis2021pixels}. Based on the successful findings of \cite{makantasis2021privileged}, we set $\epsilon=0.1$. The procedure described above results in datasets of different sizes, listed in Table \ref{table:datasize}; since regression does not use an uncertainty threshold, the entire dataset is used for both arousal and valence regression.

\begin{table}[t]
\centering
\caption{Data points for the two datasets, per time window and for classification (Class.) or regression tasks. 
}
\begin{tabular}{l||@{ }c@{ }|@{ }c@{ }|@{ }c@{ }||@{ }c@{ }|@{ }c@{ }|@{ }c@{ }}
\textbf{Datapoints} & 
\multicolumn{3}{c}{\textbf{RECOLA}} &
\multicolumn{3}{c}{\textbf{SEWA}} \\ 
\hline \hline
& 1 sec & 2 sec & 3 sec 
& 1 sec & 2 sec & 3 sec  
\\ \hline
Arousal Class. & 13752 & 13612 & 13444 & 20879 & 19590 &  18363 \\ \hline
Valence Class. & 10827 & 10750 & 10630 & 20891 & 19625 &  18512 \\ \hline
Regression & 16750 & 16692 & 16635 & 25858 & 22468 & 21150 \\ 
\end{tabular}
\label{table:datasize}
\end{table}

To evaluate the models' performance we follow a 5-fold cross-validation scheme. When splitting the dataset, we do not include data from the same participant in both training and test sets. We also use $10\%$ of the training data as a validation set to activate early stopping criteria and avoid model overfitting; training stops after 10 epochs without loss improvement on the validation set. All employed models of affect are evaluated using precisely the same data, i.e., the training, the validation and the test sets are the same for all models. Finally, for affect classification we report models' performance in terms of binary classification accuracy. When we treat affect modeling as a regression task, we evaluate models in terms of Pearson's Correlation Coefficient (PCC) \cite{benesty2009pearson} and Concordance Correlation Coefficient (CCC) \cite{lawrence1989concordance} since these metrics are widely used to quantify the performance of affect models (CCC is also used in AVEC \cite{valstar2016avec} challenges to evaluate models performance on the RECOLA dataset). PCC is used to linearly correlate two variables---in our case the predicted and ground truth affect labels---while CCC is used to measure the agreement (reliability) between the predicted and ground truth labels.

\subsection{Teacher's Impact on Student's Performance}\label{ssec:results_tuning}

\begin{table}[t]
\centering
\caption{The effect of $\alpha$ parameter on students' average binary classification accuracy (\%) when the PrivNet (top) and the FusionNet (bottom) models are used as teachers. Bold values indicate the highest classification accuracy achieved across all different values of $\alpha$.} 
\begin{tabular}{l||@{ }c@{ }|@{ }c@{ }|@{ }c@{ }||@{ }c@{ }|@{ }c@{ }|@{ }c@{ }}
\multicolumn{7}{l}{\textbf{\underline{RECOLA}}} \\ 
\textbf{PrivNet Teacher} & 
\multicolumn{3}{c}{\textbf{Arousal}} &
\multicolumn{3}{c}{\textbf{Valence}} \\ 
\hline \hline
& 1 sec & 2 sec & 3 sec 
& 1 sec & 2 sec & 3 sec  
\\ \hline
Majority Class & 50.28 & 50.39 & 50.82 & 60.75 & 60.46 & 60.99 \\ \hline
Student ($\alpha=0$)    & 60.78 & 57.34 &  61.23  & 62.32 & 62.50 & 65.48 \\ 
Student ($\alpha=0.25$)   & \textbf{64.04} & 60.29 & \textbf{62.36} & 58.08 & \textbf{64.10} & 63.72 \\ 
Student ($\alpha=0.5$)   & 60.12 & 59.21 & 59.52 & \textbf{62.69} & 63.04 & 63.14  \\
Student ($\alpha=0.75$)   & 61.87 & \textbf{64.83} & 59.01 & 62.39 & 62.09 & 63.99  \\
Student ($\alpha=1$)   & 60.25 & 60.95 & 60.84 & 62.28 & 63.28 & \textbf{65.55}  \\

\multicolumn{4}{c}{} \\

\textbf{FusionNet Teacher} & 
\multicolumn{3}{c}{\textbf{Arousal}} &
\multicolumn{3}{c}{\textbf{Valence}} \\ 
\hline \hline
& 1 sec & 2 sec & 3 sec 
& 1 sec & 2 sec & 3 sec  
\\ \hline

Majority Class & 50.28 & 50.39 & 50.82 & 60.75 & 60.46 & 60.99 \\ \hline
Student ($\alpha=0$)    & {60.78} & 57.34 & 61.23 & \textbf{62.32} & \textbf{62.50} & \textbf{65.48} \\ 
Student ($\alpha=0.25$)   & 60.14 & 59.90 & \textbf{61.34} & 60.80 & 61.69 & 60.84 \\ 
Student ($\alpha=0.5$)   & \textbf{61.06} & \textbf{61.78} & 61.06 & 61.91 & 62.05 & 63.06 \\
Student ($\alpha=0.75$)   & 59.82 & 59.28 & 56.27 & 62.22 & 60.12 & 64.04 \\
Student ($\alpha=1$)   & 60.18 & 59.78 & 58.18 & 58.56 & 58.73 & 61.80 \\

\multicolumn{7}{l}{} \\ 

\multicolumn{7}{l}{\textbf{\underline{SEWA}}} \\ 
\textbf{PrivNet Teacher} & 
\multicolumn{3}{c}{\textbf{Arousal}} &
\multicolumn{3}{c}{\textbf{Valence}} \\ 
\hline \hline
& 1 sec & 2 sec & 3 sec 
& 1 sec & 2 sec & 3 sec  
\\ \hline
Majority Class & 58.49 & 57.16  & 56.36 & 52.28 & 50.90 & 50.90 \\ \hline
Student ($\alpha=0$)    & \textbf{66.05} & 65.93  & 66.85 & \textbf{65.86} & \textbf{66.17} & 66.25\\ 
Student ($\alpha=0.25$)   & 65.73 & \textbf{66.83} & \textbf{69.10} & 65.79 & 65.46 & 67.07  \\ 
Student ($\alpha=0.5$)   & 63.64 & 66.26  & 68.13 & 65.41 & 61.43 & \textbf{67.57} \\
Student ($\alpha=0.75$)   & 61.63 & 64.26  & 63.40 & 59.10 & 63.85 & 63.32 \\
Student ($\alpha=1$)   & 60.40 & 60.72  & 62.16 & 55.85 & 60.44 & 61.71 \\

\multicolumn{4}{c}{} \\

\textbf{FusionNet Teacher} & 
\multicolumn{3}{c}{\textbf{Arousal}} &
\multicolumn{3}{c}{\textbf{Valence}} \\ 
\hline \hline
& 1 sec & 2 sec & 3 sec 
& 1 sec & 2 sec & 3 sec  
\\ \hline

Majority Class & 58.49 & 57.16  & 56.36 & 52.28 & 50.90 & 50.90 \\ \hline
Student ($\alpha=0$)    & 66.05 & 65.93  & 66.85 & \textbf{65.86} & 66.17 & 66.25\\
Student ($\alpha=0.25$)   & 64.21 & \textbf{68.58}  & 69.14 & 65.40 & 64.50 & \textbf{66.55}  \\ 
Student ($\alpha=0.5$)   & \textbf{66.62} & 66.99  & 67.97 & 63.52 & 65.86 & 66.54 \\
Student ($\alpha=0.75$)   & 66.49 & 66.83  & 68.23 & 63.52 & \textbf{66.73} & 65.26 \\
Student ($\alpha=1$)   & 65.40 & 65.39  & \textbf{70.77} & 64.94 & 64.49 & 66.04 
\\
\end{tabular}
\label{table:alpha_class}
\end{table}

We start our analysis by investigating the impact of the teacher on the performance of the student model as determined by parameter $\alpha$ in Eq.~\eqref{eq:general_loss}. We use PrivNet and FusionNet as our teacher models. Initially we train these models for modeling affect both as a classification and as a regression task. We train the student models using five values for parameter $\alpha$: $\alpha \in \{0, 0.25, 0.5, 0.75, 1\}$. With $\alpha=0$ the student considers only ground truth labels without exploiting privileged information. On the contrary, with $\alpha=1$ the student follows the teacher disregarding any information from the ground truth labels. In this study we aim to introduce the concept of privileged information in AC. Therefore, instead of using cross validation to estimate the optimal value for $\alpha$, we choose to use a set of predefined values for that parameter. We believe that presenting the performance of student models for different preset values of $\alpha$ provides better insights to the models' behaviour.
Table \ref{table:alpha_class} presents the results of this investigation. Majority class represents a classifier that always predicts the class which is the most frequently encountered in the training set.

\begin{table*}[t]
\centering
\caption{The effect of $\alpha$ parameter on students' average performance in terms of PCC/CCC when the PrivNet (top) and the FusionNet (bottom) models are used as teachers. Bold values indicate the highest PCC and CCC performance achieved across all different values of $\alpha$.}
\begin{tabular}{p{2.5cm}|>{\centering\arraybackslash}p{1.9cm}|>{\centering\arraybackslash}p{1.9cm}|>{\centering\arraybackslash}p{1.9cm}|>{\centering\arraybackslash}p{1.9cm}|>{\centering\arraybackslash}p{1.9cm}|>{\centering\arraybackslash}p{1.9cm}}

\multicolumn{7}{l}{\textbf{\underline{RECOLA}}} \\ 
\textbf{PrivNet Teacher} & \multicolumn{3}{c|}{\textbf{Arousal}} & \multicolumn{3}{c}{\textbf{Valence}} \\ \hline \hline
& 1 second & 2 seconds & 3 seconds & 1 second & 2 seconds & 3 seconds  \\ \hline
Student ($\alpha=0$)    & 0.165 / 0.100 & 0.184 / 0.095 & 0.230 / 0.166 & \textbf{0.286 / 0.166} & \textbf{0.214 / 0.133} & 0.190 / 0.138\\ 
Student ($\alpha=0.25$)   & 0.174 / 0.114 & 0.172 / 0.118 & \textbf{0.298 / 0.218} & 0.209 / 0.120 & 0.151 / 0.065 & \textbf{0.249 / 0.159} \\ 
Student ($\alpha=0.5$)   & \textbf{0.239 / 0.187} & 0.251 / 0.159 & 0.200 / 0.137 & 0.254 / 0.128 & 0.048 / 0.032 & 0.196 / 0.115\\
Student ($\alpha=0.75$)   & 0.225 / 0.118 & \textbf{0.280 / 0.179} & 0.201 / 0.132 & 0.194 / 0.105 & 0.053 / 0.047 & 0.179 / 0.097\\
Student ($\alpha=1$)   & -0.125 / -0.017 & -0.086 / -0.023 & 0.114 / 0.001 & -0.088 / 0.000 & -0.168 / -0.150 & -0.058 / -0.002\\

\multicolumn{4}{c}{} \\
\textbf{FusionNet Teacher} & \multicolumn{3}{c|}{\textbf{Arousal}} & \multicolumn{3}{c}{\textbf{Valence}} \\ \hline \hline
& 1 second & 2 seconds & 3 seconds & 1 second & 2 seconds & 3 seconds  \\ \hline
Student ($\alpha=0$)  & 0.165 / 0.100 & 0.184 / 0.095 & 0.230 / 0.166 & 0.286 / 0.166 & \textbf{0.214 / 0.133} & 0.190 / 0.138\\ 
Student ($\alpha=0.25$)   & 0.171 / 0.114 & 0.215 / 0.160 & 0.280 / 0.214 & 0.175 / 0.101 & 0.161 / 0.111 & 0.257 / 0.160\\
Student ($\alpha=0.5$)  & \textbf{0.263 / 0.169} & 0.196 / 0.111 & \textbf{0.323 / 0.235} & 0.192 / 0.112 & 0.172 / 0.111 & 0.279 / 0.162\\
Student ($\alpha=0.75$)  & 0.163 / 0.113 & \textbf{0.227 / 0.160} & 0.193 / 0.148 & \textbf{0.297 / 0.195} & 0.065 / 0.031 & \textbf{0.281 / 0.175}\\
Student ($\alpha=1$)  & 0.080 / 0.013 & -0.147 / -0.009 & 0.068 / 0.001 & -0.044 / -0.001 & -0.151 / -0.007 & -0.092 / -0.002\\

\multicolumn{7}{l}{} \\

\multicolumn{7}{l}{\textbf{\underline{SEWA}}} \\ 
\textbf{PrivNet Teacher} & \multicolumn{3}{c|}{\textbf{Arousal}} & \multicolumn{3}{c}{\textbf{Valence}} \\ \hline \hline
& 1 second & 2 seconds & 3 seconds & 1 second & 2 seconds & 3 seconds  \\ \hline
Student ($\alpha=0$) & 0.301 / 0.249 & \textbf{0.412 / 0.374} & 0.270 / 0.171 & 0.577 / 0.516 & 0.595 / 0.564 & 0.536 / 0.479\\ 
Student ($\alpha=0.25$)   & \textbf{0.399 / 0.361} & 0.393 / 0.365  & \textbf{0.391 / 0.332} & 0.589 / 0.566 & 0.611 / 0.587 & 0.499 / 0.439\\ 
Student ($\alpha=0.5$)   & 0.375 / 0.333 & 0.391 / 0.368  & 0.330 / 0.297 & \textbf{0.601 / 0.580} & 0.625 / 0.605 & \textbf{0.609 / 0.585}\\
Student ($\alpha=0.75$)   & 0.333 / 0.279 & 0.386 / 0.319  & 0.361 / 0.343 & 0.591 / 0.546 & \textbf{0.640 / 0.623} & 0.583 / 0.560\\
Student ($\alpha=1$)   & -0.077 / -0.022 & 0.006 / 0.002  & -0.132 / -0.029 & 0.082 / 0.027 & 0.054 / 0.000 & 0.115 / 0.023\\

\multicolumn{4}{c}{} \\
\textbf{FusionNet Teacher} & \multicolumn{3}{c|}{\textbf{Arousal}} & \multicolumn{3}{c}{\textbf{Valence}} \\ \hline \hline
& 1 second & 2 seconds & 3 seconds & 1 second & 2 seconds & 3 seconds  \\ \hline
Student ($\alpha=0$)    & 0.301 / 0.249 & 0.412 / 0.374 & 0.270 / 0.171 & 0.577 / 0.516 & 0.595 / 0.564 & 0.536 / 0.479\\ 
Student ($\alpha=0.25$)    & \textbf{0.388 / 0.328} & \textbf{0.418 / 0.385}  & 0.401 / 0.383 & 0.602 / 0.579 & \textbf{0.621 / 0.608} & \textbf{0.611 / 0.586}\\
Student ($\alpha=0.5$)  & 0.351 / 0.322 & 0.387 / 0.350  & \textbf{0.414 / 0.382} & 0.600 / 0.580 & 0.612 / 0.598 & 0.601 / 0.578\\
Student ($\alpha=0.75$)  & 0.363 / 0.330 & 0.383 / 0.344  & 0.294 / 0.224 & \textbf{0.611 / 0.595} & 0.621 / 0.605 & 0.593 / 0.597\\
Student ($\alpha=1$)   & 0.027 / 0.020 & -0.148 / -0.020  & 0.090 / 0.036 & -0.102 / -0.045 & -0.035 / -0.026 & -0.076 / -0.003 \\
\end{tabular}
\label{table:alpha_regr}
\end{table*}

For RECOLA, PrivNet appears to be a more powerful teacher than FusionNet. PrivNet exploits handcrafted audiovisual and physiological features, which can better capture arousal and valence \cite{ringeval2013introducing}. On the contrary, FusionNet, which uses all available modalities, seems to provide less informative predictions to the student models. This suggests, first, that privileged information can be better correlated to binary classification target variables compared to raw pixels' information and, second, that the joint distribution between raw pixels' information and target variables is not similar to the joint distribution between privileged information and target variables. Fusing modalities with dissimilar joint probability distributions will most likely deteriorate the learning model's performance since it increases the model's learning capacity (number of trainable paramters) without improving the quality of information used to model the data. As far as the affect dimensions are concerned, student models can better capture arousal than valence. For 2 seconds time windows, the best relative performance improvement between a student that exploits and the student that disregards privileged information is 13\%, while for valence the corresponding improvement is 2.5\%. This agrees with state-of-the-art results on RECOLA, which indicates that affect measurements can better capture arousal than valence \cite{zhang2018dynamic}.

For SEWA, student models seem to benefit more from the FusionNet teacher. This implies that for this dataset, raw pixels’ information and audio features complement each other. Similar to RECOLA, we observe a larger relative performance improvement for the arousal dimension (i.e. 5.8\% with three-second time windows). Interestingly, when PrivNet is used as a teacher, the student that performs best for half of the cases is the one that disregards privileged information. By combining the above two observations, we can conclude that for SEWA, the most informative modality for capturing arousal and valence is that of raw pixels.

Table \ref{table:alpha_regr} presents the results of this investigation when we treat affect modelling as a regression task.
Regarding the arousal dimension, we observe that student models’ performance improves when they exploit teachers’ information for both datasets. In this case, however, we see that the time window duration is vital for selecting the most informative teacher. For the RECOLA dataset, FusionNet appears to be a more informative teacher for 1 and 3 seconds time windows, while for SEWA, FusionNet teacher works better than PrivNet for 2 and 3 seconds time windows. Therefore, the temporal dimension highly affects the quality of information carried by each modality. 
Privileged information does not seem to improve valence modelling using 1 and 2 seconds time windows for RECOLA. In most cases, the student that disregards privileged information, irrespectively from the teacher used, achieves the best performance. For 3 seconds time windows, however, the student that exploits information from the FusionNet teacher achieves 48\% (27\%) relative performance improvement in terms of PCC (CCC) compared to the student that disregards teachers' information. For SEWA, however, information from a teacher is beneficial for the student models, irrespective of the windows' duration, resulting in a relative performance improvement of 13\% for three-second time windows with the FusionNet teacher. Similar to the classification results, FusionNet appears to be a more informative teacher.

Based on the results presented above, we can conclude that transferring knowledge from privileged information to student models can improve their performance for both regression and classification affect modeling tasks. Parameter $\alpha$ significantly affects the student models' performance and it should be appropriately set according to the problem at hand. However, forcing the student to disregard ground truth labels and follow exclusively the teacher seems to negatively affect its performance, especially for regression tasks, where we observe no or negative correlation between affect measurements and target variables. 

\subsection{The Importance of Privileged Information}\label{ssec:results_best}

In this section we investigate the impact of privileged information on building models of affect that can operate in the wild. As mentioned above, the student models make predictions using solely information that is available in the wild; in our case the raw frames of the interaction footage. We compare the student models' performance against the performance achieved by the FusionNet model that uses all information modalities captured in laboratory environments for training and testing, and PixelNet that makes predictions using the same information modalities as the student models. For the following investigation, we use non-zero $\alpha$ parameter values that yield the most accurate student models based on the sensitivity analysis covered in Section \ref{ssec:results_tuning}. Moreover, we repeat the 5-fold cross-validation scheme (see Section \ref{ssec:evaluation_framework}) five times after reshuffling the participants, to increase the robustness of statistical tests for evaluating the significance of our results. For the sake of completeness, we also present the performance of PrivNet that makes predictions using only privileged information. 

\begin{figure}[t]
	\begin{minipage}{1.0\linewidth}
		\centering
		\centerline{\includegraphics[trim={0.5cm 0.5cm 1.2cm 0.5cm},clip,width=1.0\linewidth]{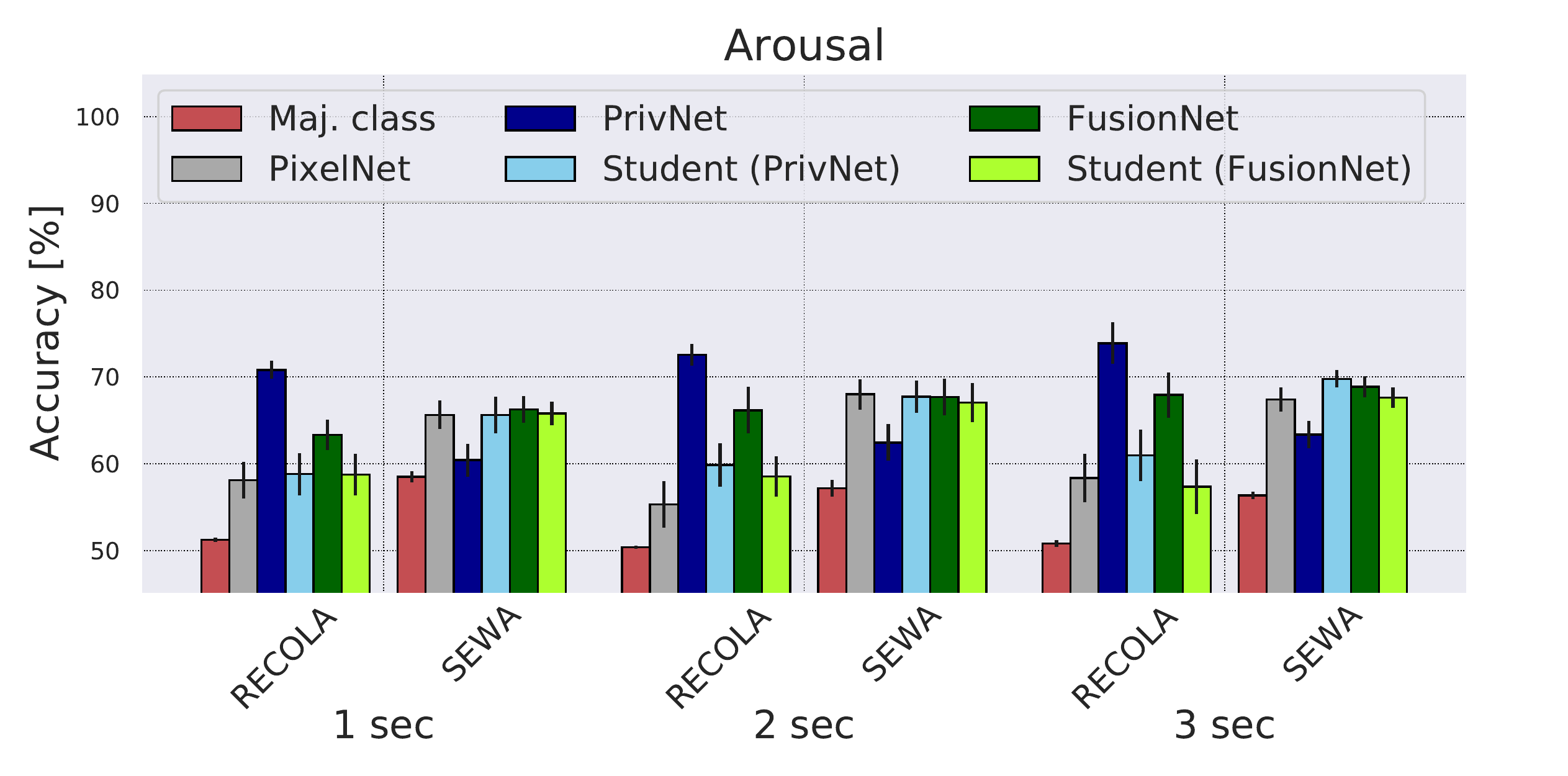}}
	\end{minipage} 
	\begin{minipage}{1.0\linewidth}
	\vspace{0.1in}
		\centering
		\centerline{\includegraphics[trim={0.5cm 0.5cm 1.2cm 0.5cm},clip,width=1.0\linewidth]{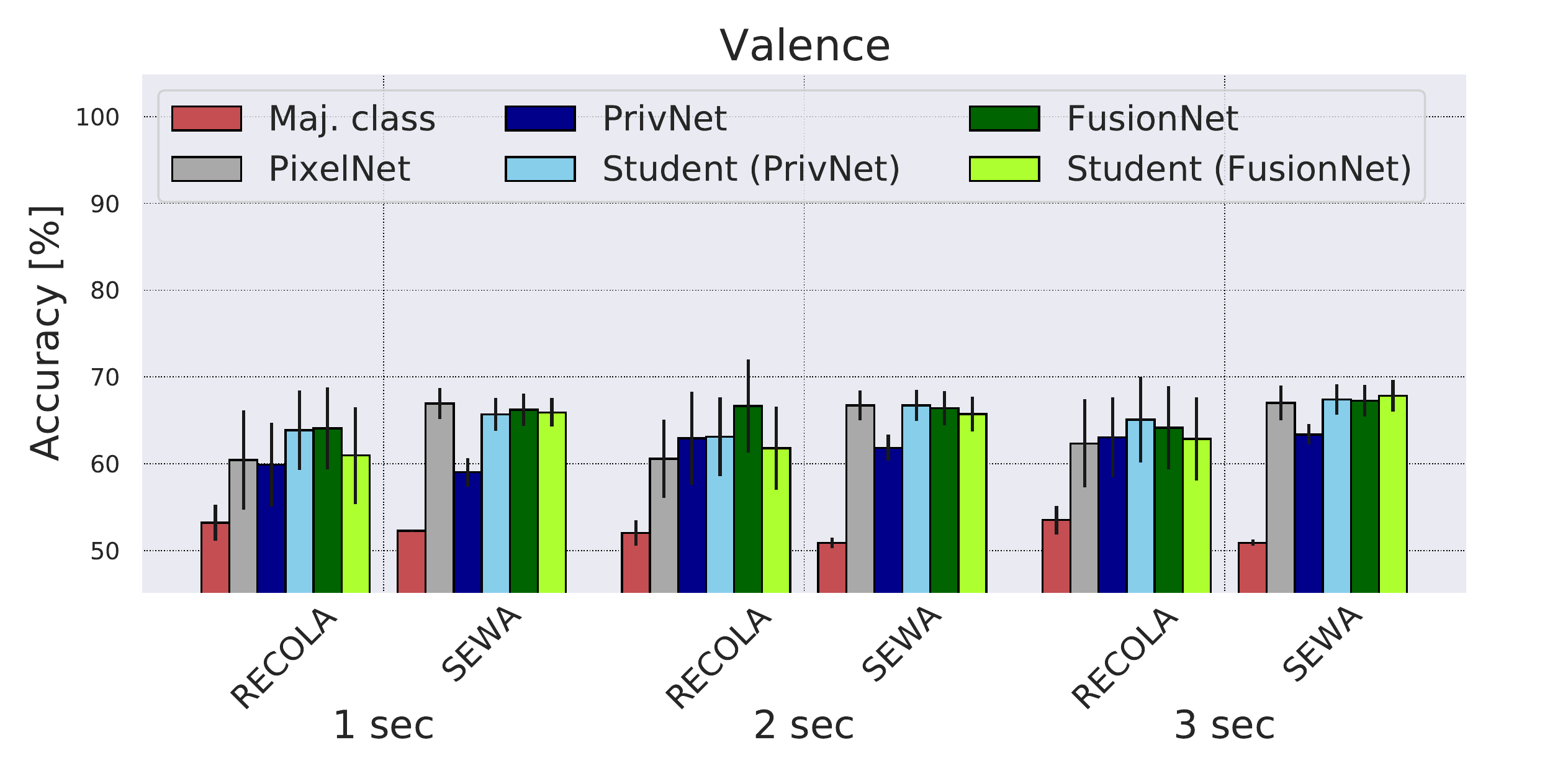}}
	\end{minipage}
	\caption{Average affect classification accuracy of different models, along with the 95\% confidence intervals across the five 5-fold cross-validation runs.}
	\label{fig:classification}
\end{figure}

Figure \ref{fig:classification} presents the results of this comparison when affect modeling is treated as a classification task. For all but one scenario (arousal classification on SEWA with 1 second time windows) and for both affect dimensions considered the student model trained using information from teachers performs on par or better than PixelNet. The relative improvement of the best student model over PixelNet is more than 4\% in 6 of 12 instances and more than 1\% in 8 of 12 instances. Out of these, one instance has over 8\% relative improvement in arousal classification using RECOLA data with 2 seconds time windows. Surprisingly, for half of the scenarios, student models perform on par with FusionNet despite the fact that the latter uses more modalities. These results suggest that student models, when appropriately parameterized, can be efficiently applied in the wild for two reasons: first, they perform on par with or better than PixelNet and second, they closely approximate the performance of FusionNet, although they use only the information that is available in the wild. The PrivNet model achieves the highest accuracy for arousal classification using the RECOLA data; this model, however, uses solely privileged information, and, thus, it cannot operate in the wild. 

\begin{figure*}[!tb]
	\begin{minipage}{0.5\linewidth}
		\centering
		\centerline{\includegraphics[trim={0.5cm 0.5cm 1.2cm 0.5cm},clip,width=1.0\linewidth]{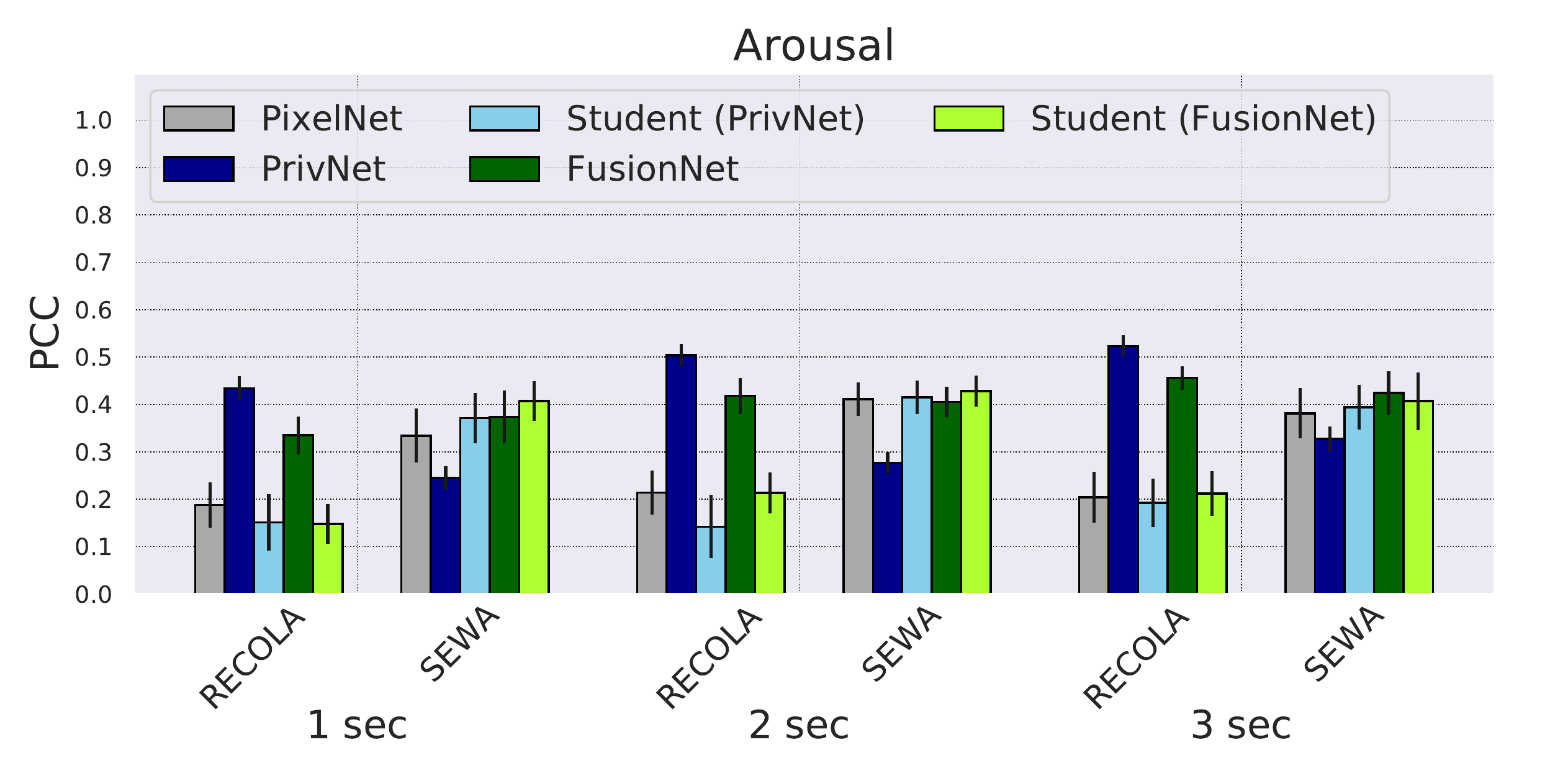}}
	\end{minipage} 
	\begin{minipage}{0.5\linewidth}
		\centering
		\centerline{\includegraphics[trim={0.5cm 0.5cm 1.2cm 0.5cm},clip,width=1.0\linewidth]{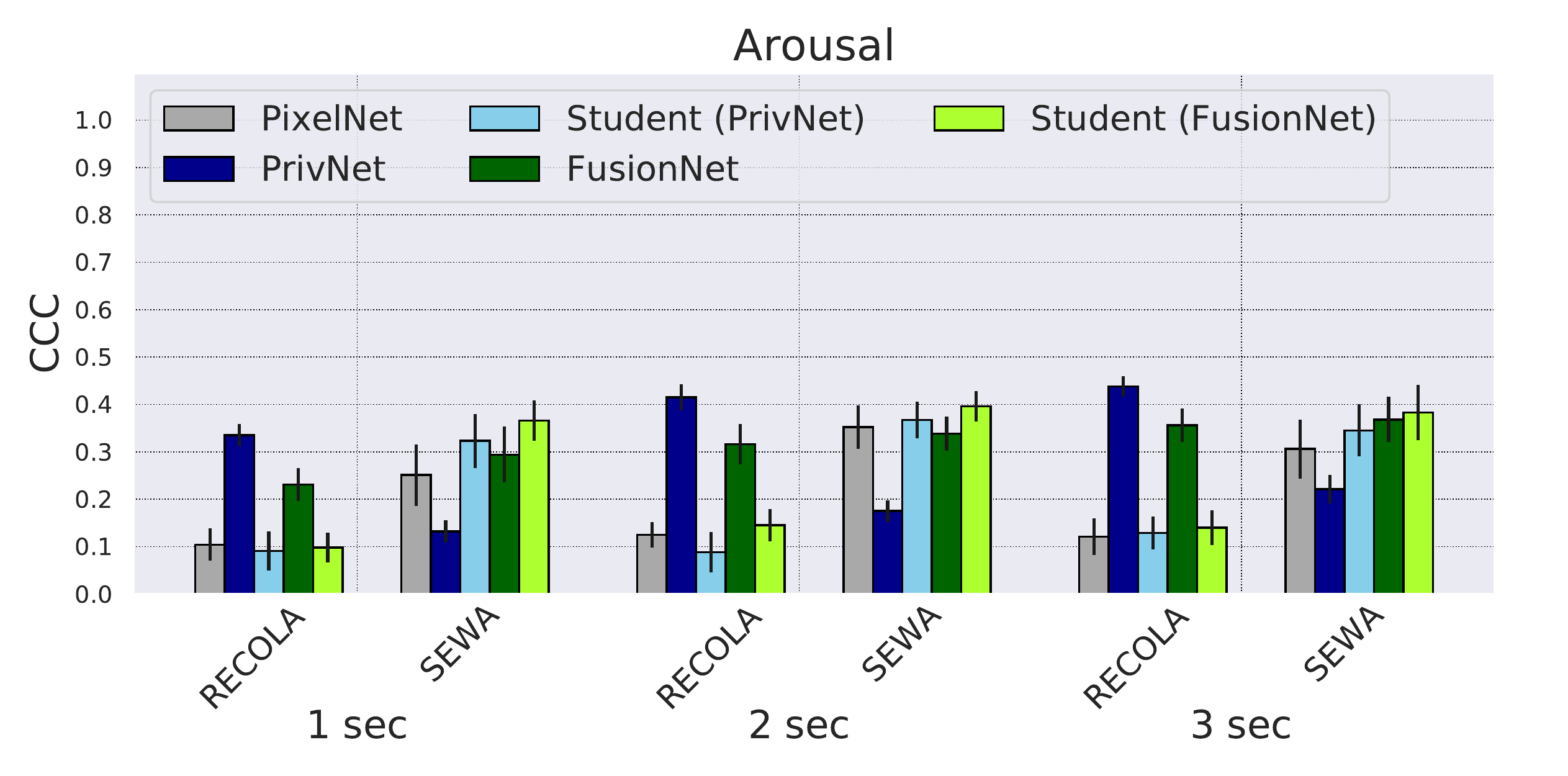}}
	\end{minipage}
	
	\begin{minipage}{0.5\linewidth}
	\vspace{0.1in}
		\centering
		\centerline{\includegraphics[trim={0.5cm 0.5cm 1.2cm 0.5cm},clip,width=1.0\linewidth]{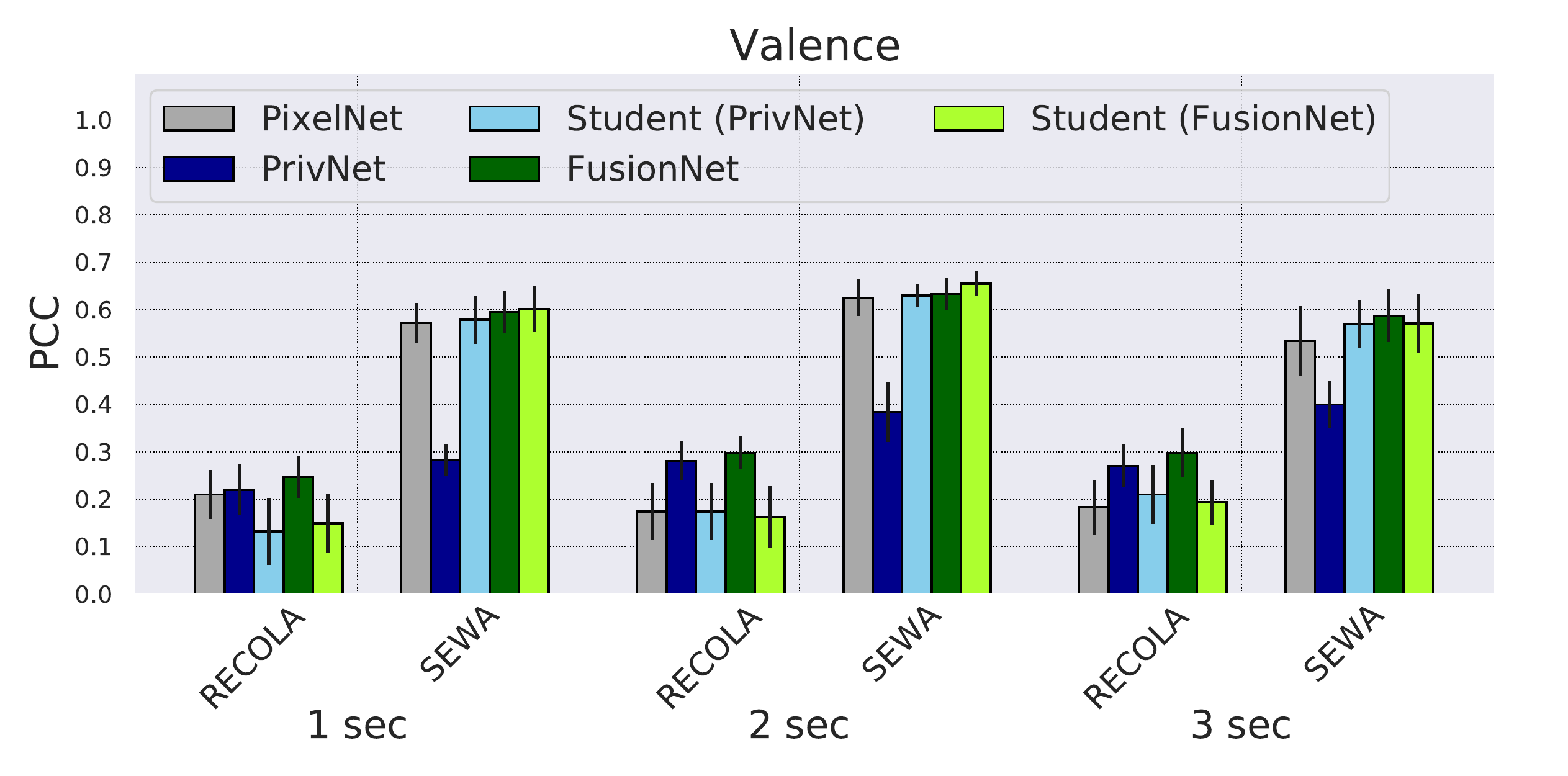}}
	\end{minipage} 
	\begin{minipage}{0.5\linewidth}
	\vspace{0.1in}
		\centering
		\centerline{\includegraphics[trim={0.5cm 0.5cm 1.2cm 0.5cm},clip,width=1.0\linewidth]{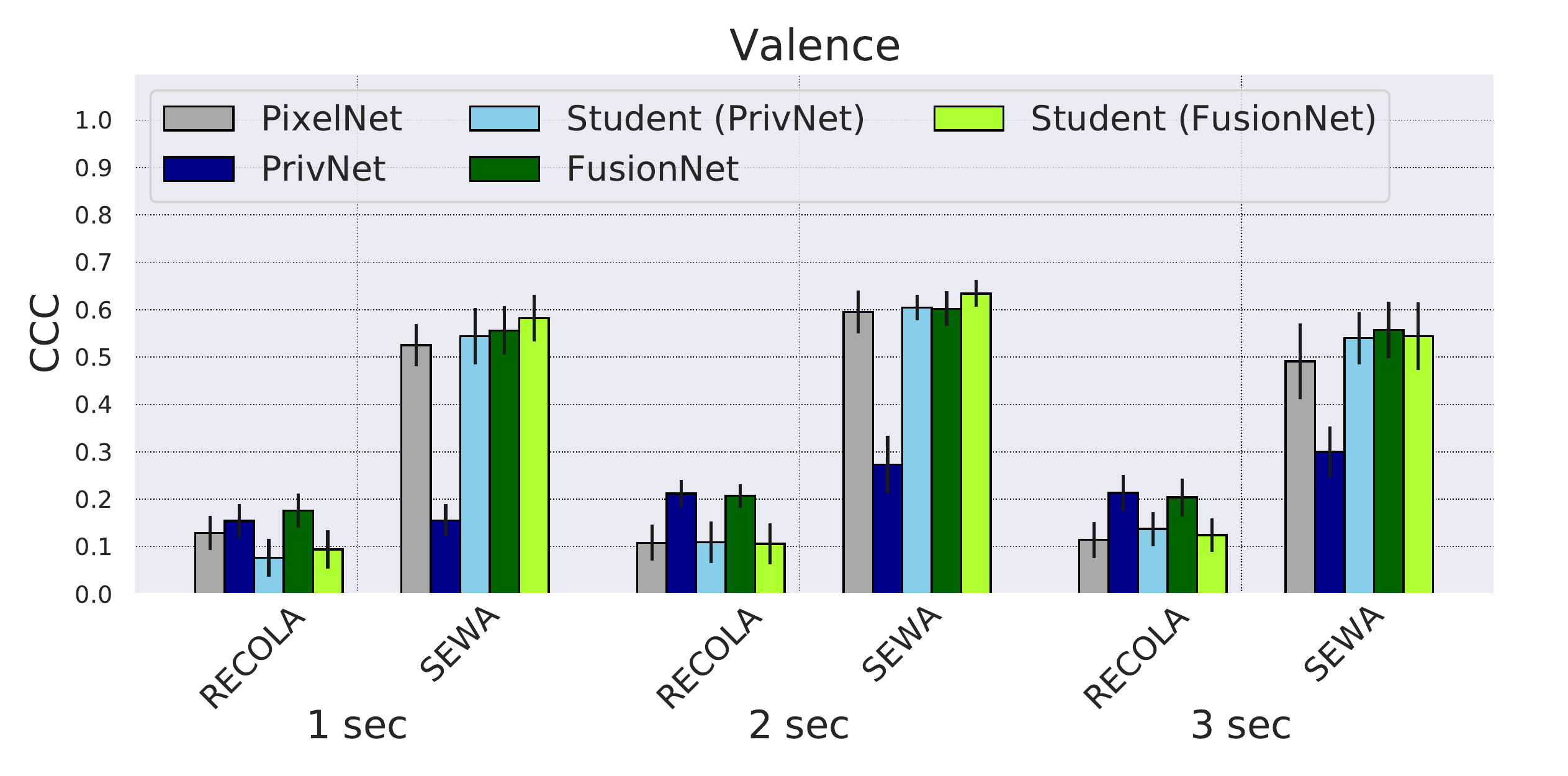}}
	\end{minipage}
	\caption{Comparison of the employed models in terms of average Pearson's Correlation Coefficient (PCC) and Concordance Correlation Coefficient (CCC) when affect modeling is treated as a regression task along with the 95\% confidence intervals across the five 5-fold cross-validation runs.}
	\label{fig:regression}
\end{figure*}

In Figure \ref{fig:regression} we compare the employed models when affect modeling is treated as a regression task. The best student model has a relative improvement over PixelNet by more than 4\% in 7 out of 12 instances for PCC and in 9 out of 12 instances for CCC. Out of these, one instance has over 45\% relative improvement in arousal regression using SEWA data with 1 second time windows. Similarly to the classification case, the performance of the student models, when appropriately parameterized, approximates the performance of FusionNet and for 9 out of 24 instances it achieves higher performance.

In both Fig. \ref{fig:classification} and Fig. \ref{fig:regression} we can also observe a large performance gap from PrivNet to the Student model that uses PrivNet as teacher. This suggests that the correlation patterns between privileged information and target variables are different from the correlation patterns between raw pixels' information and target variables. The fact that the student models try to balance between two contradicting objectives---see Eq.~\eqref{eq:class_loss}---during training likely results in the observed performance gap.

To evaluate the significance of our results, we performed the D' Agostino-Pearson normality test to check whether or not the paired differences of the tested models' performance come from a normal distribution. In the cases where the hypothesis that our data follow a normal distribution can not be rejected, we performed a one-tailed paired t-test to measure the significance of our results. However, when the normality hypothesis didn't hold, we performed the Wilcoxon signed rank test, a non-parametric version of the paired t-test. All tests were performed at a significance level of 0.05. When the RECOLA dataset is used, the student models perform significantly better than the PixelNet for 2 and 3 seconds time windows for arousal classification and for 1-second windows for valence classification. Under the regression paradigm, student models perform significantly better than PixelNet for 2 seconds time windows and 1-second time windows for arousal and valence, respectively, and for both evaluation metrics. For the SEWA dataset, the student model using PrivNet as teacher performs significantly better for 3 seconds time windows for arousal classification, and the same holds for the student model using FusionNet as teacher for 1 second time windows and valence classification. For arousal regression, student models perform significantly better than PixelNet for 1 second time windows, both for CCC and PCC. For valence regression, however, the student model that uses FusionNet as teacher performs significantly better than PixelNet, both for CCC and PCC, irrespective of the windows' length. In summary, the student model with FusionNet as teacher has significantly higher performance than PixelNet in 6 of 24 instances across datasets, while with PrivNet as teacher the student model has significantly higher performance than PixelNet in 5 of 24 instances. Overall, one or both students outperform significantly PixelNet in 8 of 24 instances, while PixelNet outperforms at least one student in 2 of 24 instances.

Based on the above experiments with two learning paradigms (classification and regression), two affect dimensions and two datasets, we conclude that student models, under the right parameters, achieve average performance values close to, or even higher, than those of FusionNet. We observe a similar behaviour across all scenarios tested, indicating that the LUPI paradigm can provide the means for building accurate models of affect that operate in the wild. Moreover, the student and the PixelNet models use the same kind of information for making predictions. The student, however, appears more robust across different scenarios and achieves on par or higher average performance for all settings. 

We should note that this study does not focus on building models that achieve state-of-the-art performance on the employed datasets. For this reason we use simple neural network architectures for our models without any parameter tuning targeting these datasets. Therefore, our results are not directly comparable to the state-of-the-art results reported for RECOLA via the AVEC challenges or for SEWA. 

\section{Discussion}

Testing affect models in the wild comes with costs associated primarily to affect sensing. One would assume that if an affect model has access to fewer modalities during testing in the wild (e.g. due to hardware/software failures or even due to the unavailability of sensors) the result will be detrimental for its accuracy. The results, however, obtained in this and our previous study \cite{makantasis2021privileged} suggest otherwise. The LUPI paradigm \cite{vapnik2009new,vapnik2015learning} provides the means to mitigate the above limitations of affect modeling in the wild. LUPI produces models that operate in the wild---having access only to a limited set of modalities (in this study raw interaction footage frames)---with small or no actual cost in performance. Our findings suggest that LUPI models can approximate or even surpass the performance of the fusion models that consider all modalities of information during both training and testing. Most importantly via LUPI affect models we gain on accessibility, cost and intrusiveness, bringing affective computing a decisive step closer to real-world applications.

In the presented case study, by examining the performance of PrivNet and PixelNet models for arousal modelling on the RECOLA dataset we see that audiovisual and physiology handcrafted features are more powerful predictors than raw footage frames. This scenario is very common in affective computing, and in general in machine learning, where task-specific handcrafted features are used to improve the performance of learning models \cite{lin2020comparison}. This fact emphasizes the importance of LUPI since student models exploit powerful handcrafted features during training to learn to make accurate predictions with low-level easy to capture information (without any dependence on the features mentioned above).

While the LUPI paradigm seems to be robust across the modalities and affect dimensions examined in this paper, our hypothesis that privileged information is beneficial for multimodal affect-based interaction needs to be tested further. Even though vanilla convolutional neural networks appear to be performing well in this and earlier studies \cite{makantasis2021pixels,makantasis2021privileged}, our plan is to test a number of different deep learning architectures for potentially improving the performance of LUPI models. Moreover, this study embeds the LUPI paradigm in the supervised affect modeling setting. We aim to extend the current methodology for transferring privileged information knowledge to semi-supervised and self-supervised learning paradigms in an attempt to learn powerful general-purpose representations for affect modeling. Another possible extension of this work is the application of LUPI beyond classification and regression to ranking and preference learning paradigms for ordinal affect modeling tasks \cite{yannakakis2018ordinal,yannakakis2017ordinal}.

\section{Conclusions}

In this paper we introduce a methodology for building models of affect in the wild exploiting \textit{privileged information}. Our hypothesis is that learning using privileged information can be used to reliably transfer affect models from controlled laboratory to uncontrolled real-world settings. To test our hypothesis we used the RECOLA and SEWA affect databases that include raw visual information, and audiovisual and physiology high-level handcrafted features. We consider all handcrafted features as privileged information (i.e. only available during model training) and assume that raw visual information corresponding to interaction footage frames is available during both training and testing. Under this setting, we treat affect modeling both as a classification and regression task, and develop models for predicting users' arousal and valence. 

The core results suggest that affect models trained using privileged information perform equally well or even better than fusion affect models that consider all modalities. Importantly for affective computing research, privileged information affect models do not require access to costly, intrusive or impractical modalities when tested in the wild. Therefore, the findings of the paper bring affective computing one step closer to realising affect interaction in the wild. The proposed methodology for knowledge transfer by following the teacher's predictions and representations has direct applications to any affect modeling task that considers multimodal data and is required to operate in the wild or to make predictions using a subset of the available modalities. Potential applications include (but are not limited to) driver-assisting systems, affective robots, affect-aware recommendation systems, affective games \cite{makantasis2021privileged,yannakakis2018artificial} and health applications at home such as stress monitoring and seizure detection.



%



\ifCLASSOPTIONcompsoc
  \section*{Acknowledgments}
\else
  \section*{Acknowledgment}
\fi
This work has been supported by the European Union’s Horizon 2020 research and innovation programme from the TAMED (Grant Agreement No. 101003397) and AI4media projects (Grant Agreement No. 951911). Antonios Liapis was supported by the Malta Council for Science and Technology (MCST) under the FUSION R\&I: Research Excellence Programme (Project number: REP-2022-017).

\ifCLASSOPTIONcaptionsoff
  \newpage
\fi



\bibliographystyle{IEEEtran}
\bibliography{IEEEabrv,affect_privileged}
%



%

\begin{IEEEbiography}[{\includegraphics[width=1in,height=1.25in,clip,keepaspectratio]{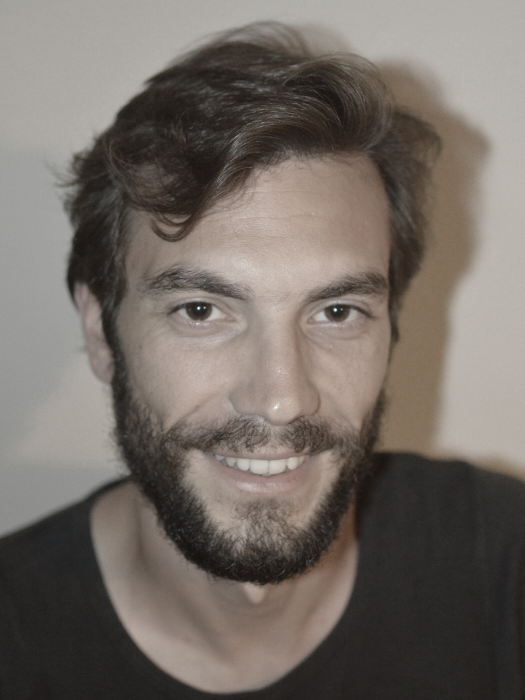}}]{Konstantinos Makantasis} is a Lecturer at the department of AI, University of Malta. He received his computer engineering diploma, his MSc and PhD from the Technical University of Crete. He has been awarded the prestigious MSCA IF Widening Fellowship to work on tensor-based machine learning methods for affect modeling. He is mostly involved and interested in computer vision, machine learning/pattern recognition, and probabilistic programming. He has more than 50 publications in international journals and conferences on computer vision, signal and image processing, and machine learning.
\end{IEEEbiography}

\begin{IEEEbiography}[{\includegraphics[width=1in,height=1.25in,clip,keepaspectratio]{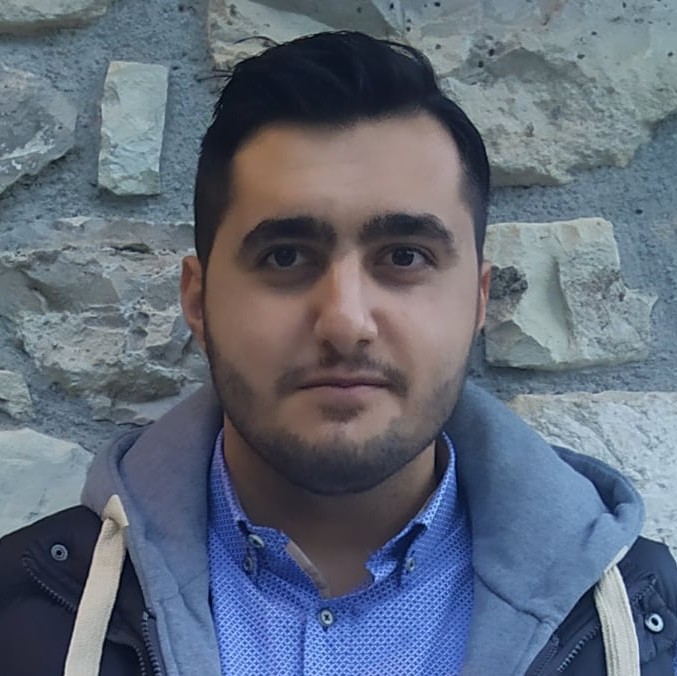}}]{Kosmas Pinitas} is a Ph.D. student at the Institute of Digital Games, University of Malta. He received his 5-year Diploma in Electrical Computer Engineering from the Technical University of Crete in 2021. His main research interests lie in the areas of preference learning, evolutionary computation, and bio-inspired learning algorithms. 
\end{IEEEbiography}
\begin{IEEEbiography}[{\includegraphics[width=1in,height=1.25in,clip,keepaspectratio]{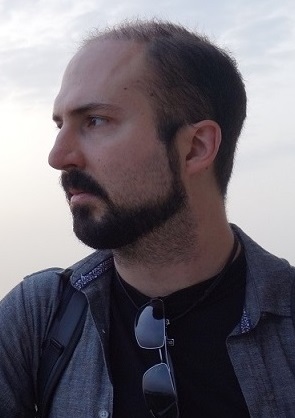}}]{Antonios Liapis} is a Senior Lecturer at the Institute of Digital Games, University of Malta. 
He received the Ph.D. degree in Information Technology from the IT University of Copenhagen in 2014. His research focuses on Artificial Intelligence in Games, Human-Computer Interaction, Computational Creativity, and User Modelling. He has published over 130 papers in the aforementioned fields, and has received several awards for his research contributions and reviewing effort. He serves as Associate Editor for the {\sc IEEE Transactions on Games}, and has served as general chair in four international conferences. 
\end{IEEEbiography}


\begin{IEEEbiography}[{\includegraphics[width=1in,height=1.25in,clip,keepaspectratio]{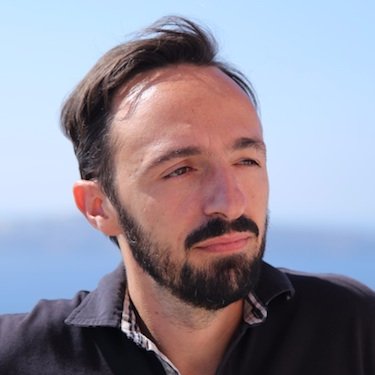}}]{Georgios N. Yannakakis}(S'04-M'05-SM'14) is a Professor and Director of the Institute of Digital Games, University of Malta, and a co-founder of modl.ai. He received the Ph.D. degree in Informatics from the University of Edinburgh in 2006. He does research at the crossroads of artificial intelligence, computational creativity, affective computing, advanced game technology, and human-computer interaction. He has published more than 300 papers in the aforementioned fields and his work has been cited broadly. He is currently the Editor in Chief of {\sc IEEE Transactions on Games} and an Associate Editor of {\sc IEEE Transactions on Evolutionary Computation}. 
Among the several awards he has received for journal and conference publications he is the recipient of the \emph{IEEE Transactions on Affective Computing Most Influential Paper Award} and the \emph{ACII 2017 Best Paper Award}. He is a senior member of the IEEE.
\end{IEEEbiography}





\vfill


\end{document}